# Would Gaze-Contingent Rendering Improve Depth Perception in Virtual and Augmented Reality?

Paul Linton[†]
Centre for Applied Vision Research
City, University of London
London, United Kingdom
paul@linton.vision

## ABSTRACT

Near distances are overestimated in virtual reality, and far distances are underestimated, but an explanation for these distortions remains elusive. One potential concern is that whilst the eye rotates to look at the virtual scene, the virtual cameras remain static. Could using eye-tracking to change the perspective of the virtual cameras as the eye rotates improve depth perception in virtual reality? This paper identifies 14 distinct perspective distortions that could in theory occur from keeping the virtual cameras fixed whilst the eye rotates in the context of near-eye displays. However, the impact of eye movements on the displayed image depends on the optical, rather than physical, distance of the display. Since the optical distance of most head-mounted displays is over 1m, most of these distortions will have only a negligible effect. The exception are 'gaze-contingent disparities', which will leave near virtual objects looking displaced from physical objects that are meant to be at the same distance in augmented reality.

## 1 Introduction

It is well-documented that near distances are overestimated in virtual reality [Rolland et al. 1995, Singh et al. 2012, Swan et al. 2015, cf. Baldassi 2015], and far distances are underestimated [Witmer and Kline 1998, Loomis and Knapp 2003, Renner et al. 2013] compared to real world performance. Over the last couple of decades a number of hypotheses have been explored, from field-of-view [Knapp and Loomis 2004, Creem-Regehr et al. 2005], to weight [Willemsen et al. 2004], to stereo-distortions [Willemsen et al. 2008], to graphics [Kunz et al. 2009], to depth of field [Langbehn et al. 2016], to vergence [Singh et al. 2012, Swan et al. 2015], to embodiment [Ries et al. 2008, Mohler et al. 2010, Leyrer et al. 2011, Lin et al. 2011, McManus et al. 2011, Creem-Regehr et al. 2015, Gonzalez-Franco et al. 2019]. None of these have fully explained the distance distortions in the literature. Instead, the recent emphasis has shifted to mitigating these distortions through training [Rousset et al. 2018], eye-height [Leyrer et al. 2015], or virtual distance offsets [Baldassi 2015].

In this paper we explore the possibility that these distortions are due to a failure to update the location of the camera in the virtual scene with the shifting gaze of the observer's eye. Our analysis proceeds in three stages. First, we outline the 14 distinct distortions of visual space that could occur when the eye rotates, but the virtual camera remains static, whilst the observer views a virtual scene through a near-eye display. These distortions are consistent with the overestimation of near distances, and the underestimation of far distances, reported in the literature. Second, we explain how these distortions of visual space can be eradicated through gaze-contingent rendering, specifically fixing the position of the camera and its frustum relative an imaginary 'window' in the virtual scene that represents the near-eye display. Third, we consider the effect that display distance and display magnification have on these distortions, and in light of which we conclude that these distortions are likely to have only a negligible effect in both CAVE environments and head-mounted displays. The exception are the 'gaze-contingent disparities', which will leave near virtual objects looking displaced from physical objects that are meant to be at the same distance in augmented reality.

## 2 Gaze-Contingent Distortions of Visual Space

Gaze contingent distortions of visual space in virtual and augmented reality could arise in one of two ways: First, the image on the display may be geometrically correct, but distorted by (a) the prismatic effect of positive lenses, and/or (b) off-axis distortions. Since these distortions are independent of the content to be rendered on the display, they would ideally be resolved through 'pre-warping' of the rendered image after it has been rendered but before it is displayed [Pohl et al. 2013]. There is no need for gaze-contingent rendering in this context. Second, the image rendered on the display may be geometrically incorrect. It is well known that rendering stereoscopic content for the wrong viewing distance or interpupillary distance will lead to distortions of visual space [Held and Banks 2008]. Could eye movements also be responsible for similar distortions in near-eye displays?

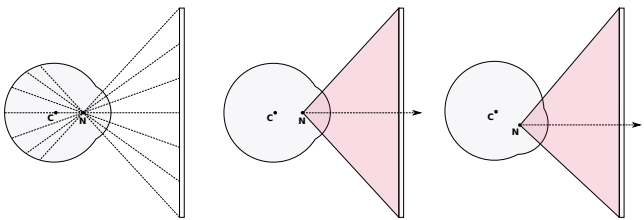

**Figure 1: Left: Human eye viewing a near-eye display. Centre: The camera frustum required to appropriately render the virtual scene. Right: The asymmetric camera frustum required to correct the distortions identified in this paper.**

The distortions we outline in this paper stem from the fact that the human eye is not rotationally centered. The left panel in Fig.1 illustrates a human eye viewing a near-eye display. Whilst the eye rotates around its centre (marked C), the light rays are actually focused through the nodal point of the eye (marked N), 6mm in front of the centre of rotation. Before we even turn to the question of eye-rotation, the first thing to notice (as illustrated by Fig.1, centre) is that to appropriately render the virtual scene we have to set the field-of-view of the camera frustum relative to the nodal point viewing the display, rather than the centre of rotation.

We now outline the 14 distortions of visual space that could occur from eye rotations in near-eye displays:

*1. Visual direction of fixated objects*: Fig.2 (left) depicts an eye rotating to fixate on a blue virtual object (the blue dot behind display), as it is presented on the display (the blue dot on display), at two time periods: pre-rotation (blue) and post-rotation (red).

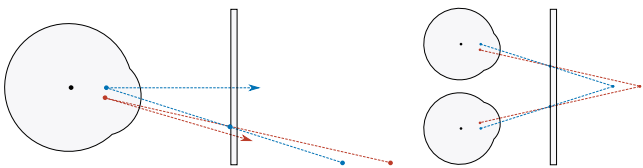

**Figure 2: Visual direction of fixated object. Left: Eye at two time periods, t1 (pre-rotation) in blue, and t2 (post-rotation) in red. Right: Resulting convergence insufficiency.**

First, let us assume that the angle of rotation is appropriately planned (red arrow). When the eye makes an eye rotation to that position (red arrow), the visual system will see that it has overshot the blue dot on the display and needs to correct (red line). This compromises the dynamics of the eye movement.

Second, once the visual axis of the eye is aligned with the object (essentially the red line), notice that the angle of rotation is shallower than it ought to be (red arrow). If height in the visual field proves to be the means by which far distances are estimated [Ooi et al. 2001, Loomis 2001], and if the visual system is able to extract this information from static fixations, then the fact that the eye adopts the red line in Fig.2 rather than the red arrow would distort distance estimates (although in the direction of distance overestimation, rather than the reported distance underestimation).

*2. Vergence insufficiency*: An equal and opposite distortion of visual direction in both eyes will lead to a shallower vergence angle (Fig.2, right), with two implications for near distance perception: First, static vergence is thought to be an absolute distance cue [Mon-Williams and Tresilian 1999, Viguier et al. 2001, cf. Linton 2017, Linton 2018], implying near distance overestimation in the right panel of Fig.2. Second, in the context of augmented reality, zero disparity (i.e. same distance) between the real and virtual objects will be achieved at the further distance specified by the red dot rather than the blue dot.

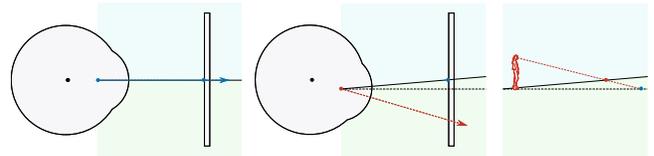

**Figure 3: Left and Centre: Eye looks downward, but horizon remains static on display (blue dot), resulting in horizon perceived as tilted up. Right: An observer viewing a tilted ground-plane views an object on the ground-plane as closer.**

*3. Visual direction of peripheral objects*: The same distortions occur for peripheral objects. We could choose any peripheral point, but the point is most clearly demonstrated with the horizon. When the eye looks downwards, the horizon becomes a peripheral object. However, the position of the horizon on the display is not updated, leading to the perception of the horizon as tilted upward. If (a) the visual angle of the horizon determines the perceived slant of the ground-plane, and if (b) the perceived slant of the ground-plane affects the distance estimates of objects on it [Li and Durgin 2012, Loomis 2014], then the observer is going to judge an object on the ground-plane to be closer than it actually is.

*4. Binocular disparity of non-fixated object*: The binocular implication of this principle is that diplopia (or double vision) of a non-fixated object, which can be a cue to distance [Morrison and Whiteside 1984], will be distorted. For instance, notice how in the right panel of Fig.4 (below), the visual direction of an object on the horizon (top and bottom red dots behind screen) are distorted into an impossible (divergent) configuration.

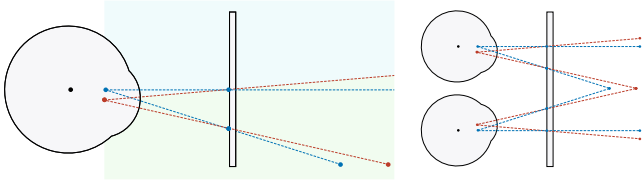

**Figure 4:** Left: Combining Fig.2 and Fig.3 illustrates how changing the rotation of the eye changes the relative visual angle. Right: An equal an opposite change in visual angle in both eyes affects relative binocular disparity.

*5. Relative visual angle*: In augmented reality, both (1) 'visual direction of fixated virtual objects', and (3) 'visual direction of peripheral virtual objects', are going to affect the relative visual angle between virtual objects and physical objects, for instance the direction of a virtual object in Fig.1 relative the true horizon.

In combination, (1) and (3) will also affect the relative direction between two virtual objects. Consider, again, an observer looking at an object on the ground-plane (Fig.4, left panel). The angle between the two blue lines is going to be slightly different than the angle between the two red lines. This angle will be increased (red) or decreased (blue) depending on whether the change in the rotation of the eye places the nodal point more perpendicular (red) or more obliquely (blue) to the centre of the two points displayed on the screen. However, this effect is likely to be small, indicating that relative distortions are likely to most pronounced in augmented rather than virtual reality.

*6. Relative binocular disparity*: An equal and opposite change in the relative visual angle in the two eyes (Fig.4, right panel) is also going to lead to a distortion of relative binocular disparities. For instance, the binocular disparity between the foreground object and an object on the horizon in Fig.4 (right) is going to be increased (red) vs. the correct binocular disparity (blue).

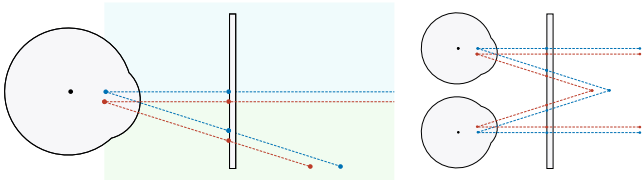

**Figure 5:** Left: 'Pre-shifting' corrects the horizon but leaves the direction of near objects distorted. Right: Implications for binocular disparity where objects seen as closer than they are.

*7. Gaze-contingent visual direction*: So far, the criticism has been that the display shows an image that would be correct for the old eye position but not the new eye position. But if eye rotations only change the position of the nodal point, you might think an easy solution is to maintain the same image, but to simply move the image up or down in the y-axis and/or left or right in the x-axis with the eye (Fig.5). This would not require gaze-contingent rendering. Instead, the image could be the 'pre-shifted' in much the same way that it is 'pre-warped' [Pohl et al. 2013].

It is true that 'pre-shifting' is an improvement. For instance, you won't get the gross error of the horizon being tilted upwards. However, this still gives us an incorrect image because the perspective of the eye changes when it moves in the real world. And the closer an object is, the more the retinal image is changed by these movements. So whilst the horizon is now correct in Fig.5 (left), the direction of the near-ground object (blue dot behind screen) is distorted to the direction of the red dot.

*8. Gaze-contingent disparities*: Equal but opposite distortions from 'pre-shifting' are also going to affect binocular disparities (Fig.5, right panel). Another way of thinking about these changes is that when the eyes converge it reduces the interpupillary distance, and this reduces the inter-nodal distance as well. If we don't reduce the binocular disparities presented on the display accordingly, then we are going to experience near objects appearing nearer than it ought to be (Fig.5, right panel). For further discussions of the gaze-contingent nature of binocular disparities see Gulick and Lawson [1976] and Turski [2016].

*9. Motion parallax of eye movements*: We have discussed static distortions of visual direction and binocular disparity. But dynamic distortions (the failure of the retinal image to change as expected) may be just as important. We can think of the movement of the nodal point of the eye as an instance of 'micro-motion parallax'. The fact that the displayed image does not update as expected when the eye moves gives us an impression of the flatness and distance of the display. This is illustrated by Fig.2 (left) and Fig.3 (left), where there is no distortion of visual direction if the virtual object is at the distance of the display. The failure of the displayed image to update with eye movements therefore acts as a counter-cue to the depth of the virtual scene.

*10. Motion parallax of head and eye movements*: Any distortions from prior eye movements will be compounded by subsequent head movements. For instance, the observer in Fig.3 will experience the slanted horizon moving forward with them.

*11. Motion parallax of head*: The same distortions will occur if we don't intentionally change our eye gaze, but only move our head. The reason being that if we are fixated on an object, whilst our head moves in one direction, our eyes will automatically counter-rotate in the other direction to maintain fixation. So, short of the gaze remaining neutral with all head movements, prioritizing head movements (where the virtual camera updates) over eye movements (where it does not) is not a solution.

*12. Proprioceptive cue-conflict*: Whenever there is a cue-conflict between visual inputs and proprioceptive information (from head and eye movements), there is likely to be visual discomfort, so it is worth considering whether a failure to update the display with eye movements contributes to the discomfort and/or motion sickness in virtual and augmented reality.

*13. Angular motion of object*: We have so far considered a static virtual world. But consider a moving virtual object being viewed by the eye. If the eye remains fixed, then the direction of motion of the object will be correctly conveyed, but as soon as the moving object is tracked by the eye, the motion of the object in terms of visual angle will be incorrect.

*14. 3D motion of object*: The same concern applies when the motion of the object is tracked by two eyes instead of one, converting the motion from visual angle motion into three-dimensional motion. If the eyes remain neutral, there will be no distortion of 3D motion, but as soon as the object is tracked by the eyes, the 3D motion of the object will be distorted.

## 3 Correcting Gaze-Contingent Distortions

Now these gaze-contingent distortions of virtual content in near-eye displays have been documented, how can they be corrected?

*1. Camera rotation*: It is tempting to think that the virtual camera should rotate with the eye. However, this is incorrect: If we model the human eye as having a constant radius between the nodal point and the retina, then the retinal image is rotationally invariant [Gillam 2007, Rogers 2007]. Because the eye is not a perfect sphere the retinal image is not rotationally invariant. But because this is a property of projection from the nodal point to the retina, so as long as we get the right projected image to the nodal point these rotational invariances in the retinal image will be produced by the eye itself as it rotates, and do not need to be artificially induced. Furthermore, so long as we (a) align the camera with the nodal point, and (b) correctly define the viewing frustum, Woods et al. [1993] illustrate that we will get the right projected image without rotating the camera.

*2. Camera location*: Instead of camera rotation, eye rotation should affect camera location: the virtual camera should move up/down and left/right with the nodal point of the eye.

*3. Viewing frustum*: On the face of it, moving the virtual camera up/down and left/right by a few millimeters would appear to be inconsequential. However, the reason these small changes in camera position cannot be ignored is that what we are trying to capture is not the absolute change in position of the nodal point, but rather its change in position relative to a fixed reference: the near-eye display, whose absolute position remains fixed.

We achieve this by replicating the near-eye display in the virtual scene as a fixed window through which the observer views the scene. And this is achieved by defining the camera frustum in the virtual scene relative to a fixed reference plane whose dimensions and absolute position in the virtual scene duplicates the dimensions and absolute position of the physical display. When coupled with the change in position of the camera, whilst not changing the camera's orientation, what this produces is an asymmetric viewing frustum of the kind proposed in Fig.1 (right).

Interestingly, Franz et al. [2004] suggest that the frustum should have a constant and static asymmetric component to account for the fact our facial structures block more of the upper visual field in the real world than the lower visual field. On our account this is a mistake, unless the display is also shifted down in the real world by the same amount to maintain the absolute relationship between frustum and near-eye display. If this is achieved, then their proposal and our proposal could be integrated by defining the frustum relative to the new display position.

## 4 Mitigating Gaze-Contingent Distortions

The asymmetric viewing frustum that we propose is the only way to eradicate gaze-contingent distortions in near-eye displays. However, these distortions can be mitigated in two ways:

*1. Display distance*: The significance of moving the camera by only a few millimeters in the virtual scene has been justified by treating the near-eye display as a fixed window through which the observer views the scene. This leads to significant changes in the viewing frustum, and therefore the image rendered on the display, when the display is close to the eye (near-eye display). However, if the fixed window is far from the observer, then millimeters changes in camera position will have virtually no effect on the displayed image. This is essentially what happens in a CAVE environment [Cruz-Neira et al. 1992] where the viewing distance is around 1.5m, so the viewing frustum will be defined relative to a fixed plane located 1.5m from the nodal point. Even if the observer looks at an object on the ground 2m away entirely using eye rotation, not head rotation, that would only lead to a displacement of the horizon by 4mm in a CAVE display (the same displacement upwards as the downwards displacement of the nodal point). Since the distance underestimation of objects on the ground-plane is still reported in CAVE environments [Marsh et al. 2014], this indicates that the gaze-contingent distortions identified in this paper cannot account for this distance underestimation.

*2. Display magnification*: In standard virtual reality headsets (Oculus Rift, HTC Vive) the display (5-6cm away) is viewed through a positive lens that both reduces the accommodative demand, and magnifies the field-of-view of the display. Field-of-view magnification is not a challenge for this account: Although the effects of eye movements are minimized relative to a magnified display, and therefore a magnified frustum, we quickly realize that the image rendered on the display will also have to be minimized by the same amount the display is magnified, thereby reinstating the 1:1 correspondence between eye movements and virtual scene content. To illustrate the point, the horizon in Fig.3 would be no less distorted if field-of-view was increased by making the physical size of the display bigger. However, magnification also increases the optical distance of the display. For instance, the magnification in Google Cardboard 2 increases the effective display distance from 5.7cm to 1.58m [Wetzstein, 2019]. This suggests that head-mounted displays ought to be no more affected by these distortions than CAVE environments.

*3. Near object disparity*: This does not mean these distortions can be completely ignored. When an observer with a 64mm interpupillary distance fixates on an object 20cm away, their inter-nodal distance effectively decreases by 2mm. If the display is located at optical infinity, this will lead to a visual angle overestimate of 0.28º in each eye. In an augmented reality environment this will correspond to the displacement of a virtual object 20cm away by 0.625cm towards the observer relative to a physical object 20cm away. Whilst this doesn't fully explain the 2-3cm displacement that Baldassi [2015] found between virtual and physical objects in near space, a 0.625cm displacement of an object may impede effective object interaction in reaching space.

# ACKNOWLEDGMENTS

We would like to thank Chris Hull for his advice on the discussion of optical magnification in Section 4.